\documentclass{iau} 
\usepackage{graphicx}
\usepackage{multicol}
\usepackage{enumitem}
\usepackage[font=small,labelfont=bf]{caption}

\title[Black hole demographics from TDE modeling] 
{Black hole demographics from TDE modeling}

\author[Mageshwaran \& Mangalam]   
{T. Mageshwaran$^1$ \and A. Mangalam$^2$}

\affiliation{Indian Institute of Astrophysics, Bangalore, India \\ email: {\tt $^{1}$mageshwaran@iiap.res.in,~~$^{2}$mangalam@iiap.res.in}}

\pubyear{2018}
\volume{342}  
\jname{Perseus in Sicily: from black hole to cluster outskirts}
\editors{Keiichi Asada, Elisabete de Gouveia dal Pino,\\ Hiroshi Nagai, Rodrigo Nemmen, Marcello Giroletti, eds.}
\begin{document}

\maketitle

\begin{abstract}

The occurence rate of tidal disruption events (TDEs) by survey missions depend on the black hole mass function of the galaxies, properties of the stellar cusp and mass of the central black hole. Using a power law density profile with Kroupa mass function, we solve the steady state Fokker-Planck to calculate the theoretical capture rate of stars by the black hole. Using a steady accretion model, the Schechter black hole mass function (BHMF) and the cosmological parameters, we calculate the detection rate of TDEs for various surveys which is then fit with the observed TDE rates to extract the Schechter parameters. The rate tension between observation ($\sim 10^{-5}~{\rm yr^{-1}}$) and theory ($\sim 10^{-4}~{\rm yr^{-1}}$) for individual galaxies is explained by the statistical average over the BHMF. 

\end{abstract}

\firstsection 
\section{Introduction}

A star orbiting close to the black hole such that the black hole's tidal gravity exceeds the star's self-gravity is tidally disrupted at the pericenter $r_p \leq r_t$, where $r_{t}= (M_{\bullet}/M_{\star})^{1/3} R_{\star}$ is the tidal radius and is called as tidal disruption events (TDEs) [Rees 1998]. The physical parameters crucial for the study of TDEs are the black hole (BH) mass $M_{\bullet}$, specific orbital energy $E$ and angular momentum $J$, star mass $M_{\star}$ and radius $R_{\star}$. For a stellar density $\rho(r) \propto r^{-\gamma}$ in the galactic center and the mass function $\xi(m)$ given by \cite[Kroupa (2001)]{Kroupa_2001}, \cite[Mageshwaran \& Mangalam (2015)]{Mageshwaran_2015} (hereafter MM15) solved the steady state Fokker-Planck equation (\cite[Merritt 2013]{Merritt_2013}) to obtain the capture rate $\dot{N}_t \propto M_{\bullet}^{-0.3}$. We use TDEs as a probe to derive the BHMF from observed detection rate for various surveys. \cite[Milosavljevi{\'c} \etal\ (2006)]{Milosa_2006} have showed that TDEs makes a negligible contribution at the higher end of luminosity functions. \cite[Stone \& Metzger (2016)]{Stone_2016} using the Schechter luminosity function showed that the volumetric rate of TDE detection is sensitive to the occupation fraction of low-mass black holes. \cite[Van Velzen (2018)]{Velzen_2018} using optical/UV selected TDEs and forward modeling showed that luminosity function $\propto L^{-2.5}$.

\vspace{-0.5cm}
\section{Black hole mass function}

We consider a separable form of BHMF $\Phi(M_{\bullet},~z)=\mu (M_{\bullet})X(z)$, where $X(z)=1-\delta(z)$ with duty cycle $\delta(z)=10^{-3}(z/0.1)^{2.5}$ [\cite[Chen \etal\ 2007]{chen_2007}] and $\mu(M_{\bullet})$ obtained using Schechter luminosity function $\Phi (L_R) = \Phi_{\star}(L_R/L_{\star})^{-\alpha}e^{-L_R/L_{\star}} $, where $L_R$ is the R-band luminosity (\cite[Schechter 1976]{Schechter_1976}). Combining the Faber-Jackson law given by $\sigma \propto L_R^{1/n}$ and $M_{\bullet}-\sigma$ relation given by $\sigma \propto M_{\bullet}^{\frac{1}{\lambda}}$, we have $\mu(M_{\bullet})=(\Phi_{\star}\epsilon/M_{s}) \left(M_{\bullet}/M_s\right)^{\beta} \exp(-(M_{\bullet}/M_s)^{\epsilon})$, where $M_s\ \propto L_{\star}^{\lambda/n}$, $\beta=\epsilon (1-\alpha)-1$ and $\epsilon=n/\lambda$. We take $n=4$ (\cite[Stone \& Metzger 2016]{Stone_2016}) and $\lambda=4.86$ (\cite[Ferrarese \& Ford 2005]{Ferrarese_2005}).

\begin{figure}[!t]
\begin{minipage}[b]{\textwidth}
  \begin{minipage}[b]{0.39\textwidth}
    \centering
\begin{tabular}{lll}
\hline
Survey & $t_{s}$ (yr) & $N_{D,obs}$  \\
\hline
ASAS-SN &  4 & 3 \\
PTF & 4 & 4 \\
iPTF & 4 & 2 \\
PS-MDS & 3.88 & 2 \\
GALEX & 10.16 & 3 \\
\hline
\end{tabular}
      \captionof{table}{The TDE surveys with the duration of survey $t_s$ and the number of TDEs detected.}
\label{tdedet}
     \end{minipage}
  \hfill
  \begin{minipage}[b]{0.59\textwidth}
    \centering
\includegraphics[width=2.5in,height=1.1in]{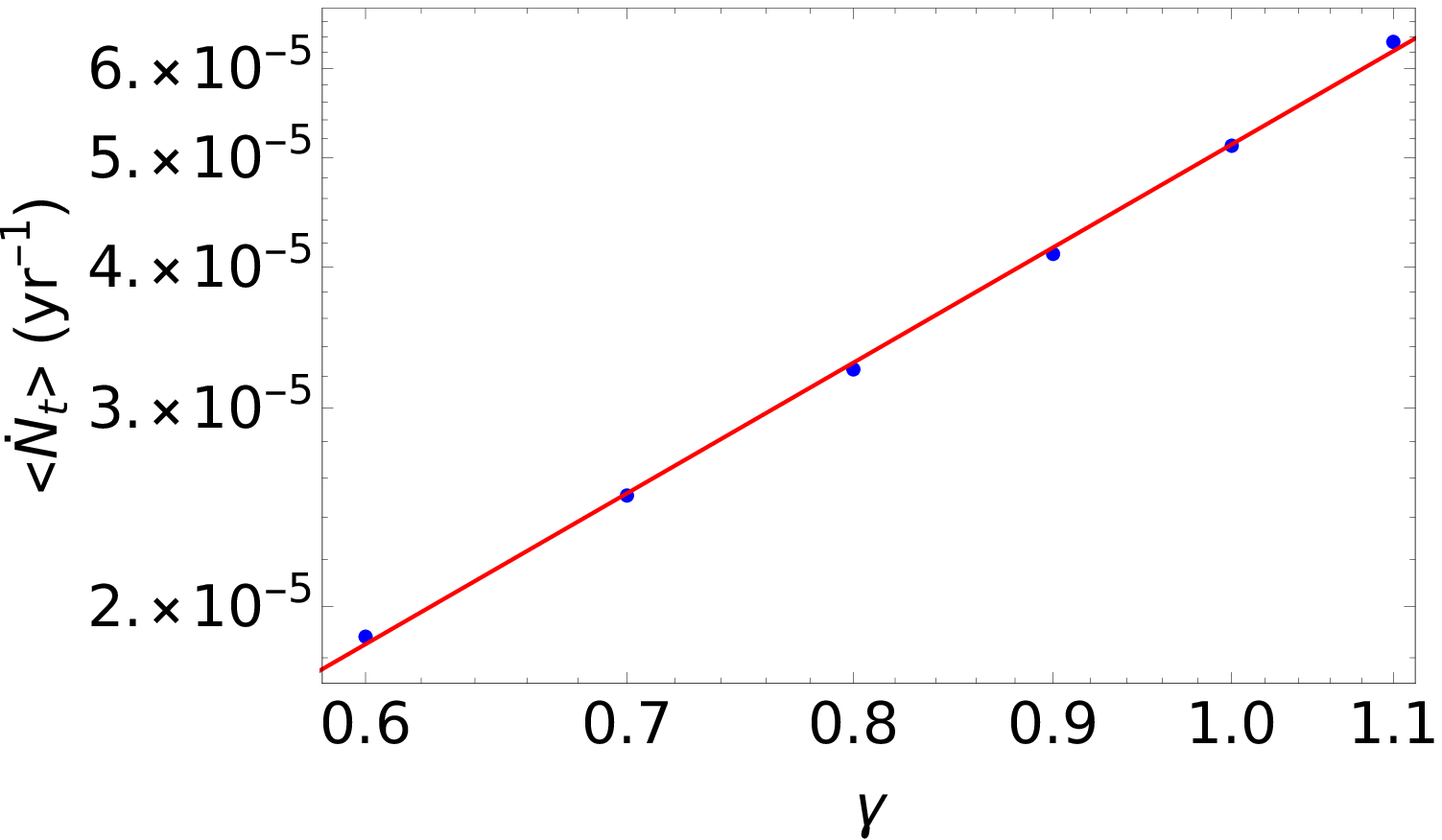}
\captionof{figure}{The galaxy averaged theoretical capture rate is shown in blue for various $\gamma$ and red line shows best fit $<\dot{N}_t> \sim 2 \times 10^{-5} \gamma^2~{\rm yr^{-1}}$.}
\label{fig1}
    \end{minipage}
  \end{minipage}
\end{figure}

The peak accretion rate and time of peak accretion for a star disrupted at $r_t$ using MM15, are given by $\dot{M}_p=7.9 \times 10^{25}~{\rm g~sec^{-1}}~ k^{\frac{3}{2}} m^{0.8} M_6^{-\frac{1}{2}}$ and $t_p=38.55 ~{\rm days}~k^{-\frac{3}{2}} m^{0.2} M_6^{\frac{1}{2}}$, where $k$ is spin-up factor taken to be 3, $M_6=M_{\bullet}/(10^6 M_{\odot})$ and $m=M_{\star}/M_{\odot}$. Assuming that the accretion rate follows the
fall back rate, the peak luminosity is given by $L_p=G M_{\bullet} \dot{M}/r_{in}=7.1 \times 10^{46}~{\rm erg~sec^{-1}}~k^{\frac{3}{2}}m^{0.8}M_6^{-\frac{1}{2}}Z^{-1}(j)$, where $r_{in}=(G M_{\bullet}/c^2)Z(j)$, $j$ is the black hole spin and $Z(j)$ is given by \cite[Bardeen \etal\ (1972)]{Bardeen_1972}. If we can calculate the $L_p$ and $t_p$ from observations, we can estimate $M_{\bullet}$ and $M_{\star}$. We consider the steady accretion disk model of MM15 for a star on a parabolic orbit with $r_p=r_t$, to calculate the peak luminosity $L_p(M_{\bullet},~M_{\star},~z)$ in the given spectral bands with the corresponding time $t_p(M_{\bullet},~M_{\star},~z)$ and assume the luminosity given by $L=L_p\left((t_p+\delta t)/t_p\right)^{-5/3}$ in the declining phase to calculate the duration of flare detection $\delta_f=t_p[\left(L_p/(4 \pi f_l d_L^2(z))\right)^{3/5}-1]$ where $f_l$ is the sensitivity of the detector. Assuming the probability of detection to be $P={\rm Min}[1,~\delta_f/(t_{cad}+t_{int})]$, where $t_{cad}$ and $t_{int}$ are the cadence and the integration times of the survey instrument, we calculate the maximum redshift $z_m (M_{\bullet},~M_{\star})$ of detection. The detection rate of TDEs for a given survey is given by \\ $\dot{N}_{D}=  4 \pi f_s d_H^3\int_{10^6 M_{\odot}}^{10^8 M_{\odot}} {\rm d} M_{\bullet} \mu (M_{\bullet})  \int_{0.8 M_{\odot}}^{150 M_{\odot}} {\rm d} M_{\star}~({\rm d} \dot{N}_{t}(\gamma,~M_{\bullet},~M_{\star})/{\rm d} M_{\star}) \int^{z_m}_{0} {\rm d} z~~  K(z)$,\\ 
where $f_s$ is fraction of sky observed, $d_H=c/H_0$, ${\rm d} \dot{N}_t/{\rm d} M_{\star}$ is obtained by integrating eqn (48) given in MM15 over $\bar{e}$ and $\ell$, $K(z)=(X(z)/(1+z))(I^2(z)/W(z))$ and $I(z)$ and $W(z)$ are given by eqn (86) in MM15. We calculate the number of detections in survey time $t_s$ for the surveys given in Table \ref{tdedet} and compare it with the observed rate to calculate the Schechter parameters.

\vspace{-0.5cm}
\section{Results}

We found that $\gamma=0.9$, is the most possible solution to match the observed Schechter parameters and the best inferred Schechter parameters from our model fit is $\alpha=1.093$, $\phi_{\star}=10^{-4}~{\rm Mpc^{-3}}~{\rm and}~ L_{\star}=1.13 \times 10^{42}~{\rm erg~sec^{-1}}$. The galaxy averaged capture rate calculated using derived BHMF is $<\dot{N}_t> \sim 2 \times 10^{-5} \gamma^2~{\rm yr^{-1}}$ as shown in Fig \ref{fig1} which is comparable with observed TDE rates $\sim 10^{-5}~{\rm yr^{-1}}$(\cite[Donley \etal\ 2002]{Donley_etal2002}). The estimation of BHMF from TDEs will improve with larger sample sizes which will be possible in the near future with the upcoming surveys like LSST, eROSITA, and zPTF.
\setlength{\columnsep}{1cm}
\vspace{-0.5cm}
\bibliographystyle{unsrt}

\begin{thebibliography}{99}
\def\baselinestretch{0.7}  
\vspace{-0.4cm}
\scriptsize
\begin{multicols}{2}
\bibitem[Bardeen, Press, \& Teukolsky (1972)]{Bardeen_1972}
{Bardeen, J.~M., Press, W.~H.,  \& Teukolsky, S.~A.} 1972,
\textit{ApJ},178, 347

\bibitem[Chen, Wang \& Zhang (2007)]{Hopkins_2007}
{Chen, Y.-M., Wang, J.-M.,  \& Zhang, F.} 2007,
\textit{in Astronomical Society of the Pacific Conference Series}, Vol. 373, 667

\bibitem[Donley \etal\ (2002)]{Donley_etal2002}
{Donley, J.~L., Brandt, W.~N., Eracleous, M.,  \& Boller, T} 1995,
\textit{AJ}, 124, 1308

\bibitem[Ferrarese \& Ford (2005)]{Ferrarese_2005}
{Ferrarese, L.,  \& Ford, H.} 2005,
\textit{Space Science Reviews}, 116, 523
\bibitem[Kroupa (2001)]{Kroupa_2001}
{Kroupa, P.} 2001,
\textit{MNRAS}, 322, 231
\columnbreak
\bibitem[Mageshwaran \& Mangalam (2015)]{Mageshwaran_2015}
{Mageshwaran, T.,  \& Mangalam, A} 2015,
\textit{ApJ}, 814, 141

\bibitem[Merritt (2013)]{Merritt_2013}
{Merritt, D} 2013, 
\textit{Dynamics and Evolution of Galactic Nuclei} (Princeton University Press)
\bibitem[Milosavljevi{\'c}, Merritt \& Ho (2006)]{Milosa_2006}
{Milosavljevi{\'c}, M., Merritt, D.,  \& Ho, L.~C.} 2006,
\textit{ApJ}, 652, 120
\bibitem[Schechter (1976)]{Schechter_1976}
{Schechter, P.} 1976,
\textit{ApJ}, 203, 297
\bibitem[Stone \& Metzger (2016)]{Stone_2016}
{Stone, N.~C.,  \& Metzger, B.~D.} 2016,
\textit{MNRAS}, 455, 859
\bibitem[Van Velzen (2018)]{Velzen_2018}
{van Velzen, S.} 2018,
\textit{ApJ}, 852, 72
\end{multicols}
\end{thebibliography}

\end{document}